\documentclass[acmsmall]{acmart}

\setcopyright{acmlicensed}
\acmJournal{PACMHCI}
\acmYear{2024} 
\acmVolume{8} 
\acmNumber{CSCW1} 
\acmArticle{208} 
\acmMonth{4}
\acmDOI{10.1145/3653699}

\usepackage{csquotes}
\usepackage{ifdraft}
\usepackage{etoolbox} 
\usepackage{xspace}
\usepackage{color}
\usepackage{tabularray}
\usepackage{comment}
\usepackage{cleveref}
\renewcommand*{\cref}{\Cref}
\usepackage{hyphenat}
\usepackage{array}
\usepackage{booktabs}
\usepackage{tabularx} 
\usepackage{pgfplots}
\usepackage{multirow}
\usepackage{subcaption}
\AtBeginDocument{%
  \providecommand\BibTeX{{%
    \normalfont B\kern-0.5em{\scshape i\kern-0.25em b}\kern-0.8em\TeX}}}

\AtBeginDocument{
  \providecommand\BibTeX{{
    \normalfont B\kern-0.5em{\scshape i\kern-0.25em b}\kern-0.8em\TeX}}}

\NewDocumentCommand{\pquote}{+O{} +m}{\textit{\blockquote[#1]{#2}}}

\newcommand{\elide}[1]{\textelp{}} 
 
\newcommand{\textcite}[1]{\citeauthor{#1}~\cite{#1}}

\newcommand{\change}[1]{\textcolor{black}{#1}}

\received{January 2023}
\received[revised]{October 2023}
\received[accepted]{January 2024}

\begin{document}

\title{Modeling Health Video Consumption Behaviors on Social Media: Activities, Challenges, and Characteristics}

\author{Jiaying Liu}
\email{jiayingliu@utexas.edu}
\orcid{0000-0002-5398-1485 }
\affiliation{
  \institution{School of Information, The University of Texas at Austin}
  \country{USA}
 }

\author{Yan Zhang}
\email{yanz@utexas.edu}
\orcid{0000-0002-1130-0012}
\affiliation{
  \institution{School of Information, The University of Texas at Austin}
  \country{USA}
}


\renewcommand{\shortauthors}{Jiaying Liu \& Yan Zhang}

\begin{abstract}
    Many people now watch \textit{health videos} on various topics, such as diet, exercise, mental health, COVID-19, and chronic disease videos, on social media. Most existing studies focused on video creators, leaving the motivations and practices of viewers underexplored. We interviewed 18 participants and surveyed 121 respondents and derived a model characterizing consumers' video consumption practices on social media. The practices include five main activities: \textit{deciding} to watch videos driven by various motivations, \textit{accessing} videos on social media through a socio-technical ecosystem across different types of platforms, \textit{watching} videos to meet informational, emotional, and entertainment needs, \textit{evaluating} the credibility and interestingness of videos, and \textit{using} videos to achieve health goals. Through an iterative video consumption process, individuals strategically navigate across multiple platforms, seeking better accessibility, higher reliability, and cultivating a stronger motivation. They actively look for longer and more in-depth videos. We further identified challenges consumers face while consuming health videos on social media and discussed design implications and directions for future research.
\end{abstract}



\begin{teaserfigure}
  \centering
  \includegraphics[width= .9\textwidth]{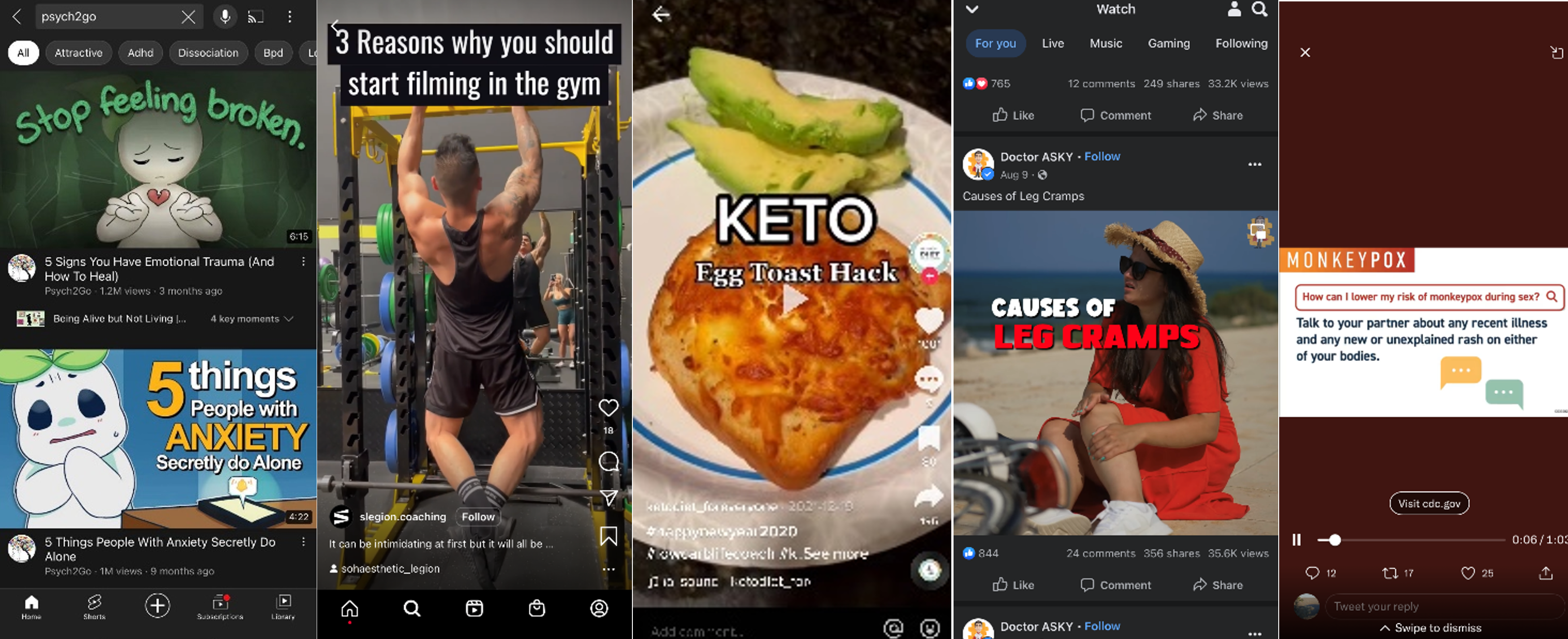}
  \caption{Examples of health videos on social media (from left to right: a mental health video on YouTube, a Gymhack exercise video on Instagram, a keto diet video on TikTok, a physical injury video on Facebook, and a monkeypox video on \change{X (previous Twitter)})}
  \Description{Examples of health videos on social media (from left to right: a mental health video on YouTube, a Gymhack exercise video on Instagram, a keto diet video on TikTok, a physical injury video on Facebook, a Monkeypox video on \change{X (previous Twitter)}}
  \label{fig:videos}
\end{teaserfigure}

\maketitle

\section{Introduction}
Consuming videos on social media is increasingly popular. A typical user now spends two hours and twenty-five minutes on social media each day \cite{digital_2021_digital_2021}, and 49\% of social media users watch more than five videos per day \cite{mccue_state_2020}. Around 3.37 billion people consumed videos in 2022, accounting for 91.8\% of Internet users. 
Consumers use multiple platforms for video consumption. 
TikTok, YouTube, and Twitch are the three largest video-sharing platforms \cite{bartolome2023literature} with rich video content. Platforms that previously focused on text and pictures rolled out sections designated for videos. Now, users spend more than half the time on Instagram Reels and Facebook Watch watching videos \cite{the_wall_street_journal_metas_2022}, 82\% \change{X (previous Twitter)} users watch videos, and Snapchat users consume 10 billion videos every day \cite{villa_10_2021}.

Health videos are popular on social media, particularly after the COVID-19 pandemic. 
\textit{Health videos} refer to online videos about health-related topics, such as mental health, sexual health, and physical exercise. Fig ~\ref{fig:videos} shows some examples. Unlike live-streaming, online videos are not synchronized, thus could afford more convenience and attract a larger audience \cite{rigby2018can}.

Health videos on social media have a great potential to influence consumers' knowledge, health decisions, and behaviors. Early evidence has shown that health videos may have contributed to the exacerbation of eating disorders \cite{jordan_can_2021} and inaccurate self-diagnoses of mental disorders \cite{jargon_tiktok_2021}. 
Despite the impact on consumers, much of the existing research on health videos has concentrated on understanding video creators' motivations, such as disease management \cite{huh_health_2014} and community building \cite{liu_health_2013}, with limited attention given to consumer practices. Gaining a thorough understanding of consumers' experiences with health videos is vital to guide the effective use of videos and social media to promote positive health behaviors.

\change{Another goal of this study is to contribute to a conceptual understanding of video consumption behaviors. Video consumption is intrinsically connected with various fields within HCI and CSCW. Researchers have explored video consumption in contexts including learning \cite{lee_personalizing_2021}, online dating \cite{he2023seeking}, and e-commerce \cite{tolunay2021analyzing}. However, these studies primarily centered on video consumption on a specific platform; thus, a holistic understanding of user behaviors in the complex social and technical environment is still needed.  Theoretically, prior studies only revealed a subset of viewers' video consumption behaviors, such as viewers' watching motivations or their perception of videos' characteristics \cite{lee_personalizing_2021}. Therefore, it is worth investigating how to model and conceptualize consumers' overall video consumption process.}

Our study addresses these gaps by exploring consumers' lived experiences of health video consumption in the complex socio-technical environment consisting of various digital platforms and social relationships. We attempt to provide a conceptual understanding of users' practices and activities in relation to health video consumption.
Our research questions are:
\begin{itemize}
\item RQ1: What are consumers' practices of consuming health videos on social media?
\item RQ2: What challenges do consumers face in consuming health videos on social media?
\end{itemize}

We conducted a mixed-method study where we interviewed 18 participants with diverse demographics and varying health goals and surveyed 121 respondents. Our participants reported their experiences with watching health videos on a wide range of topics, such as general wellness (e.g., diet and exercise), chronic diseases (e.g., obesity and diabetes), and mental health. Based on the analysis of the interviews, we identified five major activities involved in consuming health videos on social media, including deciding, accessing, watching, evaluating, and using. 

Participants \textit{decide} to watch health videos with motivations for health improvement, emotional support, learning knowledge, social awareness, and entertainment. They \textit{access} health videos across topics, devices, platforms, and channels through searching, browsing, and social recommendations. When \textit{watching} health videos, the video modality enables participants to acquire information with greater clarity and efficiency, emotionally engage with the content, and explore various genres. Participants \textit{evaluate} health videos based on the platform previews, content, and creators. Finally, participants \textit{use} health videos in four ways: guiding problem-solving, personalizing long-term behavior change solutions, regulating emotions alone, and managing video content. 
It should be noted that health video consumption behaviors are dynamic; thus, the five activities are not linear; rather, they are \textit{iterative} in nature. For example, the evaluation and use of health videos might lead to changes in consumption motivations and access platforms. Specifically, we observed several prominent behavioral trends in the iterative consumption of health videos, including developing stronger motivation, seeking better accessibility, and navigating to in-depth and reliable content.
Further, we summarized the challenges that participants encountered at each activity and described how the barriers cascade.

This work is a timely response to the surge of health video consumption on social media. We contribute to research on online videos and consumer health behaviors by (1) demonstrating that health videos on social media are becoming an important health information source that consumers actively use to satisfy multiple needs; (2) depicting the cross-platform and iterative consumption behaviors within the socio-technical ecosystem and indicating several crucial health video consumption preferences; (3) identifying challenges of health videos consumption and proposing corresponding design implications; and (4) providing a preliminary five-activity model of health video consumption behaviors on social media, which could potentially characterize user video consumption practices beyond the health context. 

\section{Related Work}
We reviewed prior research on online video research and specifically health videos in HCI and identified several research gaps on the topic.

\subsection{Online videos consumption behaviors}
Online videos are gaining popularity \cite{bentley2019exploring}, and video consumption behavior is now ubiquitous on a number of platforms, including TV sites, social media platforms, and news websites \cite{bentley2019understanding}. Online videos provide ample resources to serve on-demand and cross-device watching at varying times and locations \cite{rigby2018can}. 

A large stream of research in CSCW concentrated on live video streaming in various contexts such as gaming \cite{lessel_expanding_2017, sjoblom_why_2017}, learning \cite{lee_personalizing_2021}, cooking \cite{tang_meerkat_2016}, eating \cite{anjani_why_2020}, and traveling \cite{haimson_what_2017}. Studies suggested that the \enquote{live} nature of video streaming is key to attracting and engaging viewers by having real-time conversations with the audience to enhance interactivity.
However, according to statistics, live streaming video consumption only accounts for 28.5\% \cite{shepherd_30_nodate} of the Internet video traffic, and online video consumption is more omnipresent because it offers the convenience of flexible viewing and attracts a larger audience \cite{lu_i_2019}. Thus, this study focused on the consumption of online videos to address the lack of relevant studies.

Previous studies reported some features of online video regarding user engagement.
Although online videos are prepackaged and asynchronized, they still create para-social interactions (PSI) \cite{horton_mass_1956}. For example, \textcite{niu_stayhome_2021} summarized six social provisions of \#WithMe videos, including attachment, integration, reassurance, alliance, guidance, and nurturance. These provisions facilitate and foster contextualized disclosure, creator-viewer interactions, and community-building. 
Creators' techniques such as talking directly to the camera \cite{niu_stayhome_2021} also play vital roles in cultivating a sense of intimacy and connectedness \cite{anjani_why_2020}.
Platform functions enable viewers to respond to and interact directly with creators by commenting and liking \cite{wohn_explaining_2018}. 

\change{Despite health topics constituting a large part of online videos, little is known about why and how people engage with health videos. To build on prior work on more play-oriented or entertainment-related topics like gaming and traveling, we focused on how videos are used in the topic domain of health.} 

\subsection{Health videos on social media}
Social media play a vital role in people’s health information acquisition \cite{carrotte2015predictors, goodyear_young_2019} and health videos now pervade social media platforms \cite{harris_mixed_2019}. Major public health agencies, such as the CDC, and many health non-profit organizations maintain video channels on popular social media platforms such as YouTube and Facebook for public health education and awareness raising. User-generated health videos that share illness experiences, disease management, or lifestyle videos further enrich the diversity of health content. During the COVID-19 pandemic, videos about health topics received even more attention and served as resources for COVID-19 news \cite{li_communicating_2021}, companies to mitigate loneliness \cite{niu_stayhome_2021}, windows to peer into doctors’ life \cite{southerton2021research}, and facilitators to keep fit at home \cite{ibrahim_virtual_2021}. 

However, the existing research mainly focused on video creators' motivations to produce heath videos on social media, such as managing chronic diseases \cite{huh_health_2014}, building a supportive community \cite{liu_health_2013}, disseminating educational materials \cite{poquet_video_2018}, keeping a personal diary for fighting diseases \cite{noyes2004video}, and expanding visibility on social media \cite{sakib_does_2020}. Researchers comprehensively surveyed the challenges encountered by creators during video production and proposed design implications to support their needs of engaging viewers \cite{fraser_sharing_2019}, managing fans, improving workflow efficiency \cite{weber_its_2021}, and promoting e-commerce activities \cite{haimson_what_2017}. 

Only a limited number of studies explored viewers' motivations for consuming health videos on social media. For example, \textcite{harris_mixed_2019} found young people use YouTube videos to learn about a wide range of health topics. \textcite{song_interventions_2021} surveyed why people watch health videos on TikTok and identified three significant contributing affordances– social presence, immersion, and credibility perception. \change{However, prior studies primarily focused on a specific activity in video consumption, such as motivations or the evaluation of health videos. A holistic understanding of the range of activities involved in the video consumption process is needed. Further, these studies mostly examined a specific health topic (e.g., unhealthy food \cite{coates_its_2020} and sexual behavior \cite{levinson_qualitative_2020}) on a single platform (e.g., TikTok or YouTube). Thus, they focused less on the complexity and dynamics of the lived experiences of consumers' health video consumption on the socio-technical ecosystem of social media.}

The pressing issue of low-quality health videos underscores the urgency of delving into people's experiences and challenges when engaging with such content. Health videos on social media are increasingly used as decision aids for the general public \cite{haslam2019youtube}. However, their quality remains a significant concern. Various studies invited health professionals to assess the quality of health videos, including those related to diabetes on TikTok \cite{kong2021tiktok}, meningioma treatment on YouTube \cite{sledzinska_quality_2021}, and cancer on Instagram \cite{xu2021instagram}. These investigations revealed substantial and alarming differences in the quality of health videos. 

Overall, the current studies are characterized by topic-specific, platform-specific, and activity-specific inquiries. We aimed to extend the existing research by holistically studying why and how people consume health videos on social media.

\subsection{Models of content consumption on social media}
Another goal of the study is to achieve an improved conceptual understanding of consumers' video consumption behavior on social media. 
A number of models and theories have been widely used to explain user behaviors on social media. 
One of the most frequently used theories is the Use and Gratification Theory (UGT) \cite{katz1973uses}, which has been validated by much empirical research and constantly amended and renewed in response to the changing media economy \cite{ruggiero2000uses}. It is presumed that people use social media to fulfill certain needs. 

\textcite{kaplan_users_2010} specifically classified two types of behaviors on social media: content consumption and self-disclosure/ social interactions. 
Based on this, some HCI researchers conducted empirical research and developed frameworks explaining self-disclosure behaviors on social media in health-related contexts, such as LGBTQ identity \cite{devito_too_2018} and chronic illnesses \cite{shamon2019attention}. For example, \textcite{pyle_lgbtq_2021} provided a framework of why LGBTQ+ people decide to disclose pregnancy loss on social media. 
However, there are no models that capture how content consumption unfolds on social media. 

Some information seeking and retrieval models can, to some degree, be used to make sense of the content consumption process on social media. For example, the berry-picking model \cite{bates1989design} proposes that people's real-life information search process is like picking berries in bushes where the information seeker learns from documents retrieved and moves from one source to the other to meet information needs. The schematic model of information seeking \cite{savolainen2006time} assumes that information seeking begins with information needs, which lead them to identify and access information sources, then examine and evaluate information, followed by using information. 

\change{However, these models tend to focus on active information search and pay less attention to content consumption through browsing and encountering, which are common on social media. Additionally, these models are developed and applied mostly in studies that examine people's search for text-based information but less in studies on video consumption where the unique features and affordances of video modality need to be considered.} A model that is specific to video consumption on social media is needed to formalize our study results to guide future research on video consumption in the context of social media.

\section{Methods}
We used the mixed-method approach, incorporating semi-structured interviews and surveys with two different sets of participants. The interviews allowed us to elicit people's narratives about their lived experience of consuming health videos on social media, while the survey provided a general view of video consumption behaviors and prevented us from losing sight of a broad picture when interpreting the interviews.
Specifically, we designed the survey after we conducted and analyzed the first 12 interviews. The analysis of survey results further informed the following theoretical sampling of new interviewees.
The research protocols were approved by the University of Texas at Austin Institutional Review Board. 

\subsection{The semi-structured interview}
We began with interviews to reveal contextualized and detailed consumers' motivations, practices, and interactions with health videos on social media.

\subsubsection{Recruitment}
Previous literature indicated that socioeconomic backgrounds significantly influence people's health behaviors \cite{schaefbauer_snack_2015}. Thus, in addition to our university listserv, we recruited participants from Reddit to bring in hard-to-reach and low-economic-status participants \cite{topolovec2016use}. The included participants showed a wide diversity and confirmed our recruitment efforts.

Interested participants completed a screening questionnaire reporting demographics (e.g., age, gender, race, health condition) and health video consumption experiences (e.g., topics, motivations, frequency, and platforms). Responses from social media that were likely generated by bots were removed \cite{chen2023integrating}. Further, we adopted the theoretical sampling strategy \cite{charmaz2000grounded}, where we intentionally chose participants to maximize the sample demographic diversity and diversity in video-viewing behaviors. For example, we carefully selected participants with different health goals and included those interested in wellness topics and relatively severe conditions such as depression. We also included participants watching health videos for their own needs and family or friends' needs.

\subsubsection{Participants}
In total, 18 participants were interviewed (9 females and 9 males). Their age ranged from 18 to 80, with the median age being 29. Most of them (16/18) reported having mild or serious health concerns or conditions, ranging from BMI control to depression and Parkinson's. All reported using health videos on social media to fulfill their health goals. Table ~\ref{tab: participants} shows more details about the participants. 

\begin{table}[h]
\centering
\caption{Participants Characteristics}
\label{tab: participants}
\resizebox{\textwidth}{!}{
\begin{tabular}{lllllllll}
    \toprule
ID  & Sex & Age & Race & Education     & Frequency & Platforms                & Health   status             & Health   goal \\
    \midrule
P01 & M      & 39 & Asian
 & Graduate degree     & Monthly   & YT                       & Chalazion                       & Unachieved  \\
P02 & F      & 21 & Black & Undergraduate     & Daily     & TT,   YT                 & Obesity                         & Achieved    \\
P03 & F      & 20 & Asian & Undergraduate     & Daily     & TT,   YT, IN, RD         & Knee pain, vegan lifestyle              & Achieved   \\
P04 & F      & 20  & Asian & Undergraduate     & Daily     & TT,   YT, IN             & Obesity                         & Unachieved  \\
P05 & M      & 21  &White & Undergraduate   & Daily     & YT,   RD                 & Depression                      & Unachieved  \\
P06 & M      & 21  & Asian & Undergraduate    & Daily     & TT,   YT, TW, FB, IN, RD & No condition                         & Achieved    \\
P07 &
  F &
  24 &  Asian &
  Graduate student &
  Daily &
  TT,   YT, TW, FB, IN &
  BMI   control, father's high blood pressure &
  Achieved \\
P08 & M      & 25 & Black & High school    & Daily     & TT,   YT, TW, FB         & Depression                      & Achieved    \\
P09 &
  F &
  25 & Hispanic &
  Graduate student & Daily  &
  TT,   YT, TW, FB, IN &
  Friends' asthma &
  Achieved \\
P10 & F      & 29 &Hispanic & Graduate degree    & Weekly    & YT,   FB, IN, RD         & No condition                            & Achieved    \\
P11 & M      & 18 & Multiple Races & Undergraduate & Daily     & TT,   YT, IN, RD         & Prediabetes                     & Achieved    \\
P12 & F      & 18 & White & Undergraduate    & Daily     & TT,   YT, IN             & Obesity                         & Unachieved  \\
P13 & M      & 26  & Hispanic & Undergraduate & Weekly    & YT                       & Asthma                          & Achieved    \\
  \multirow{2}{*}{P14} &
  \multirow{2}{*}{F} &
  \multirow{2}{*}{46} &
\multirow{2}{*}{White} &
  \multirow{2}{*}{Graduate student} &
  \multirow{2}{*}{Daily} &
  \multirow{2}{*}{YT, IN} &
  ADHD &
  Unachieved \\
    &        &     &          &      &     &                          & Mobility   issue                & Achieved    \\
P15 & M      & 43 & Black & Graduate student    & Daily     & YT,   IN                 & Obesity,   uncle's heart attack & Achieved    \\
P16 & M      & 35 &White & High school    & Weekly    & YT                       & ADHD,   EDS                     & Achieved    \\
P17 & M      & 42 & Black & High school    & Daily     & YT,   TT                 & Obesity,   grandma's diabetes   & Achieved    \\
P18 & F      & 80 & White & Bachelor degree & Weekly    & YT                       & Parkinson                       & Achieved   \\
  \bottomrule
\end{tabular}
}\\
\scriptsize Note: In the column of platforms, YT=YouTube, TT=TikTok, IN=Instagram, FB=Facebook, TW=Twitter, RD=Reddit. In the Health status column, BMI= Body Mass Index, ADHD= Attention-deficit/hyperactivity disorder, and EDS= Ehlers-Danlos syndrome. 

\end{table}

\subsubsection{Interview procedure}
After getting participants' consent, the first author interviewed all participants on Zoom, a video conferencing platform, because of the quarantine during the COVID-19 pandemic. All interviews were recorded and transcribed using Zoom. The interviews lasted, on average, 1.5 hours. Each participant received a \$20 Amazon gift card at the end of the interview. 

Each interview began with grand tour questions asking about participants' general social media use and health video consumption practices. Example questions include \enquote{In general, on which social media platforms do you watch health videos? What topics do you usually watch? Why do you watch these topics?}

With such background information in mind, then, to enhance data specificity and quality, we used the critical incident \cite{flanagan_critical_1954} and the contextual inquiry techniques \cite{beyer1999contextual}. We informed the participants that \enquote{We want to learn more examples about how and why you watched health videos on the platforms you mentioned earlier.} To help participants recall more cases, we started by asking about the most recent or most impressive video they watched and elicited more detailed follow-up questions such as \enquote{How did you find this video? Did you think this video helpful?} We utilized questions regarding this incident to familiarize participants with the depth and scope of our interview and then further prompted experiences with other incidents. The interviewer went over every topic and platform mentioned by each participant. Example questions include, \enquote{Did you remember the first time you watched health videos [on a certain platform/ about a certain topic]? Why did you stop/continue watching such videos? After that, what other health videos do you watch?} We also asked about the most satisfying and unsatisfying health videos they watched. We evoked the contexts in these narratives to understand why they watched and interacted with videos, how the health videos influenced them, how they navigated multiple platforms, and what challenges they encountered.

The last part of the interview asked the participants to compare their experiences with health information in different modalities (e.g., video, text, radio, face-to-face) and on different platforms.

\subsubsection{Data analysis and Model development}
The interview data were analyzed using the iterative open coding and axial coding method \cite{corbin_basics_2014}. 

The first author wrote memos immediately after each interview and during the analysis process. The analysis started when we completed the first five interviews so emergent themes could be compared with previous literature and further explored in subsequent interviews. For example, several participants reported that accessing health videos often involved multiple platforms, and their preference for platform changed after watching and comparing videos across platforms; the theme of the iterative process was created to represent this practice, and the theme was further explored and developed in subsequent interviews. The first author developed an initial open coding schema based on the five interviews that included themes and sub-themes related to participants’ practices and motivations for consuming health videos on social media, as well as the challenges they encountered.

This schema was constantly revised in weekly discussions between the two authors based on the first author's notes and debriefs of the interviews conducted in the past week and the comparison of these interviews with prior interviews. The two authors proceeded to axial coding by revisiting the open codes and the corresponding transcripts to refine and conceptualize reoccurring themes into categories. These codes were stabilized after the 12th interview. For example, different motivations for watching health videos were integrated into the deciding activity category. Five categories that represented activities during the video consumption process emerged from the analysis, including motivations, access channels, videos' advantages during watching, evaluation, and using videos.

We revisited the interviews and mapped the five categories according to the chronological order of participants' narratives. We revised the names of the categories and built an initial model. This model was further developed in the following interviews and analysis process. For example, the evaluation and use activities were not adequately represented in the previous 12th interviews; therefore, we selectively chose participants with behavioral change goals and focused on eliciting their accounts of evaluating and using videos for the following interviews. We also compared the activities with prior empirical studies and models such as the Use and Gratification Theory \cite{katz1973uses} and the schematic model of information seeking \cite{savolainen2006time}.

\subsection{The Survey}
The survey was conducted to reveal a broader picture of consumers' health video consumption practices to enrich the perspectives obtained from the interviews.

\subsubsection{Participants}
We collected survey responses from Amazon Mechanical Turk (AMT) and Reddit. Studies indicated that the AMT pool has a similar demographic distribution as the overall US population \cite{kittur_crowdsourcing_2008}. 
We retained 121 responses from the 548 responses after excluding incomplete or irrelevant responses \cite{chen2023integrating}. Among the respondents, 54 (44.6\%) were female, 63 (52.1\%) were male, three (4.5\%) were non-binary/third gender, and one (0.8\%) preferred not to respond. Their age ranged from 18-76, with the median age being 29. Regarding education, 46 (38\%) had high school/ vocational training degrees or less, 61 (50.4\%) had Bachelor's degrees, and 14 (11.6\%) had Graduate degrees.

\subsubsection{The survey instrument}
The survey contained 33 questions and was launched on Qualtrics. The survey instrument asked about respondents' demographic information and video consumption practices on social media. We designed a series of multiple-choice questions based on the categories and themes identified in the qualitative analysis. For example, the options for the question of motivation came from subthemes that represented participants' specific goals. All multiple-choice questions included \enquote{others, please specify:} to gather other types of activities. We attempted to validate and develop these new activities in the following interviews.
To enhance data quality, we added one attention check question \cite{august_pay_2019, shamon2019attention} and improved the wording and visual design of the survey with two rounds of tests. The appendix \ref{suveydetails} includes more details about the survey design.

\subsubsection{Data analysis}
We used the survey responses to corroborate and compare with the qualitative findings. We conducted a descriptive analysis of survey respondents’ health video consumption practices. The results shed light on the prevalence of the five categories and their subcategories, namely, specific activities of health video consumption. 

\subsection{Reflections on the research design}
Utilizing virtual interviews turned out to be a strength of our study. Interviewing participants via Zoom enabled us to reach out to low-resource communities who can not afford the time and transportation for in-person meetings. Among the participants, two had never used Zoom but were able to complete the interviews successfully with the researchers' assistance.

Nevertheless, our research design also has limitations. First, our recruitment of survey and interview respondents only included people who have watched health videos on social media while excluding those without such experiences. 
Second, user-generated content (UGC) was inadvertently overrepresented in our interviews. Most participants watched individual-created videos, while very few participants accessed videos from non-profit organizations and government agencies. This may limit our understanding of their access and evaluating practices. 
Third, videos about more severe health conditions, such as cancer, were not covered in the study, such that we could have missed some health video consumption motivations and practices. 
Fourth, participants' descriptions of experiences with health videos might deviate from their real behaviors, especially for instances that happened years ago. 
Fifth, collecting data via participants' recall might overrepresent experiences on YouTube because they may have forgotten many short health videos that they randomly watched without a strong motivation on other platforms. 

\section{RQ1: Modeling Consumer Health Video Consumption Practices on social media}
\begin{figure}[h]
  \centering
  \includegraphics[width=\linewidth]{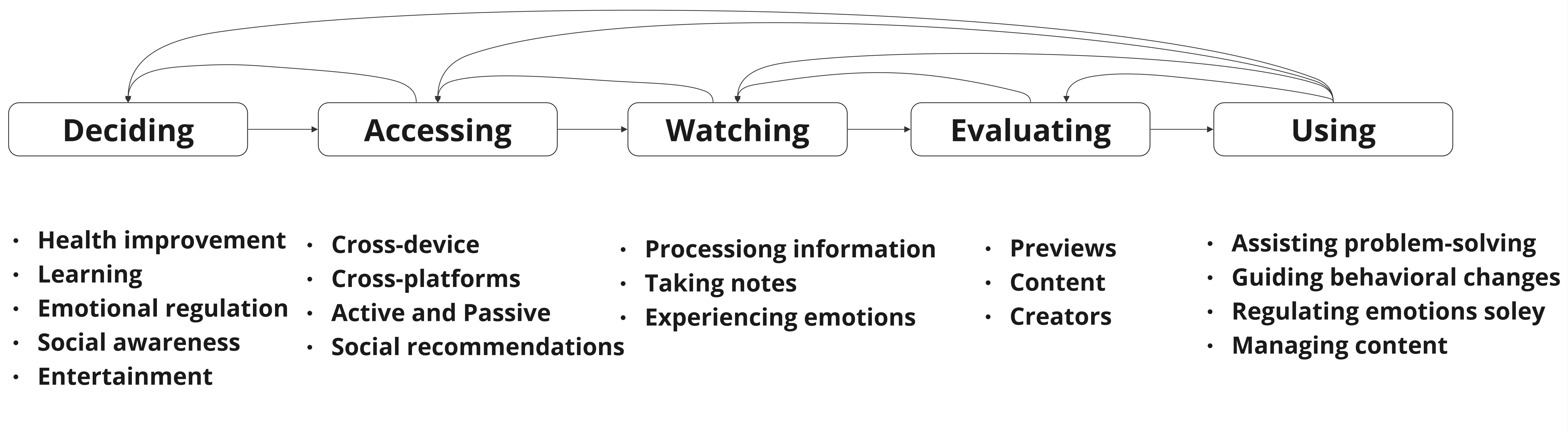}
  \caption{A Model of Video Consumption on Social Media. This model consists of five main activities: \textit{deciding} to watch videos by certain motivations, \textit{accessing} health videos through a socio-technical ecosystem, \textit{watching} videos by engaging with information and emotions, \textit{evaluating} videos by features of access channels, content, and creators, and \textit{using} videos to influence behavior, knowledge, and emotions. These activities do not necessarily proceed in a linear fashion; consumers might \textbf{skip} or \textbf{return} to previous activities in an \textbf{iterative} way.}
  \Description{A model of video consumption on social media. This model consists of five main activities: \textit{deciding} to watch videos by certain motivations, \textit{accessing} health videos through a socio-technical ecosystem, \textit{watching} videos by engaging with information and emotions, \textit{evaluating} videos by features of access channels, content, and creators, and \textit{using} videos to influence behavior, knowledge, and emotions. These activities do not necessarily proceed in a linear fashion; consumers might \textbf{skip} or \textbf{return} to previous activities when performing one activity in an \textbf{iterative} way.}
  \label{fig:model}
\end{figure}

The model, shown in Fig \ref{fig:model}, portrays that consumers' health video consumption on social media consists of five main activities. 

\subsection{Deciding to watch health videos to fulfill motivations}
\begin{figure}[h]
  \centering
  \includegraphics[width=0.8\linewidth]{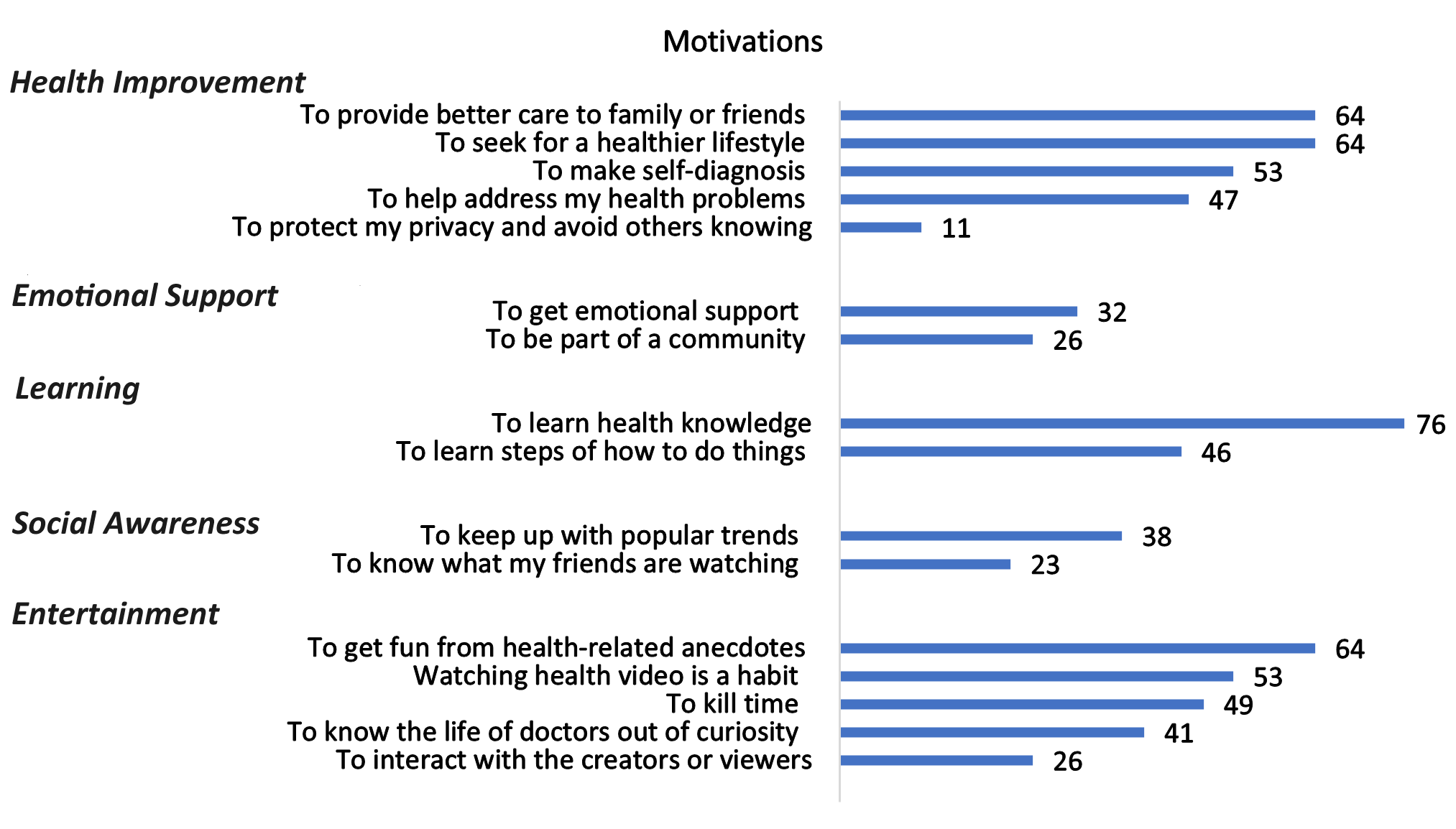}
  \caption{Survey results of motivations for consuming health videos on social media}
  \Description{Survey results of motivations for watching health videos on social media}
  \label{fig: motivations}
\end{figure}

Deciding to watch health videos is often driven by specific motivations. Five major categories of motivations were identified: health improvement, emotional support, learning, social awareness, and entertainment. 
Figure \ref{fig: motivations} shows the survey results concerning each category of motivations, demonstrating the prevalence of the motivations among a larger sample than the interview participants. 

\subsubsection{Health Improvement}
Health improvement includes motivations to develop healthier lifestyles as a preventative measure, address existing health concerns, and provide care to family and friends. This was a major motivation for consuming health videos.

Many people are proactive in preventing diseases. For example, P05 said “\textit{I want to make sure I'm informed of mask usage, and I was almost constantly waiting for videos about the fact. I'm not really at risk, but I don't like being sick and suffering.}” 

Under circumstances when formal medical treatment fails, is unaffordable, or conditions are stigmatized, the accessibility of health videos on social media motivates viewers to use them as a substitute for healthcare services. For example, P05 suffered severe depression but did not turn to any family members, friends, or professionals for help, because “\textit{I just think that mental health was not something that was very open for discussion in my background.}” P08 relied only on videos produced by a YouTuber to cope with his depression because he could not afford professional consultation.

Health video consumption was not limited to serving personal usage; it also served as a resource to care for family and friends. For example, P14 had a WhatsApp family group where he and his brother shared health videos weekly or monthly because “\textit{we have family members from other states and countries. Sharing the YouTube links to the group is the best way I tried to provide healthcare to make sure that they're aware that good health can be beneficial to themselves.}” P07 and P09 reported searching for videos about high blood pressure and asthma respectively in order to take care of diagnosed family members and hospitalized friends.

\subsubsection{Emotional support}
Seeking emotional support was another major motivation for consuming health videos, particularly for a sense of community, hope, and self-regulation. 

Sense of community was described as feelings that \pquote[P10]{you're not the only person}. Participants wanted to find people who shared similar health conditions or health beliefs. For example, P16 had Ehlers-Danlos syndrome (EDS), a rare disease, which took him a long time to be officially diagnosed and pressured him both physically and mentally. He described “\textit{it was quite reassuring to me to find people talking about how long it took them to get a diagnosis.}” Another participant, P05, watched many commentary videos that criticize anti-vaxxers because “\textit{I'm glad I'm not the only one who thinks that's what the people [anti-vaxxers] are saying is kind of crazy.}” 

Hope can emotionally support many participants to overcome challenges. P02 commented on health videos on YouTube that “\textit{Maybe I'm struggling with something right now, but there are people who have gone through it and they are offering solutions. The fact that I can have access to these videos makes me a more hopeful person.}” P16 was also inspired by videos about “\textit{real stories on YouTube, examples of people who have lived with overweight}” so that he “\textit{saw the challenges they have gone through, but they managed to survive.}”

Encouragement offers the emotional incentive to make efforts. Encouragement is essential for them to maintain healthy activities, such as exercising regularly and eating healthy. They went to TikTok and Instagram for motivational health videos. For example, P03 explained that “\textit{watching these motivates me to get out of bed and get working. I see other people doing things that are healthy, and they're working out, so to motivate me.}” 

\subsubsection{Learning}
All participants reported learning knowledge being one of the motives for watching health videos. Much of such learning was informal with a focus on problem-solving. However, several participants were serious about learning knowledge from health videos and performed more structured and sustained learning to build one’s knowledge repertoire on particular topics. For example, P02 had a “\textit{Tactical Tuesday},” a time to watch videos that “\textit{are focused on my mind, trying to build my brain, and developing emotional intelligence because that's something that you can continuously build.}” P13 watched health videos on YouTube almost every day “\textit{because they are something not only new but also very useful. Maybe not for yourself, maybe not for today or tomorrow, but you never know in the future or someone else.}” P01 also kept watching videos of conference presentations, TED talks, and lectures to acquire knowledge about psychology.

\subsubsection{Social awareness}
Social awareness represents the urge to keep up with recent social events and social relationships. “\textit{Trends}” and friends' interests were mentioned as a motivation frequently. For example, P04 mentioned her attention to an exercise campaign on TikTok because “\textit{This is a really huge thing during the Covid-19 like all my friends and every girl is doing that}.” Mental health was another common content our participants encountered on Facebook and Instagram. Many perceived that “\textit{Mental health is a huge thing recently.}” During National Bullying Month, P07 watched mental health videos because “\textit{there are at least two friends shared [on Facebook]... I saw some friends commented and added a new link, being like ‘Hey, check it out, this is also good stuff’. Then I just take a quick look at it.}” She watched not only the shared videos but also a few other related videos recommended by YouTube. 

\subsubsection{Entertainment}
Entertainment is another important motivation for watching health videos. Several participants watched amusing videos, such as exercise challenges and emergency room anecdotes, on Instagram, TikTok, and YouTube. For example, P05 watched commentary videos where the influencer played videos containing anti-vaccine viewpoints and rebutted their claims because “\textit{he senses things very humorously and his responses are very strong.}” P06 subscribed to health workers’ channels on YouTube and Instagram and watched videos about hospital anecdotes to relax. P10 mentioned that she watched Doctor Mike, a YouTube influencer who made many funny health videos during breaks and lunches, for entertainment.

\subsection{Accessing videos}
\begin{figure}[h]
  \centering
  \includegraphics[width=0.8\linewidth]{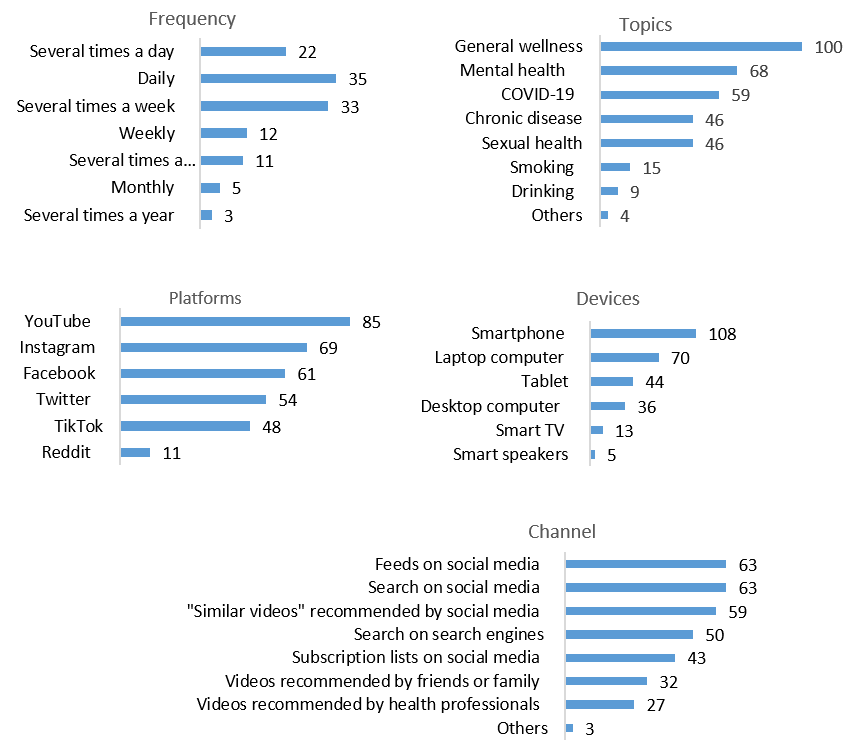}
  \caption{Basic aspects of consumer video accessing activity on social media}
  \Description{Bar charts of devices, topics, frequencies, platforms, and access methods of health video-watching practices. Nearly 84\% of the survey respondents watched health videos on social media at least one time a week. General health and wellness topics (e.g., exercise, diet, sleep) were watched the most frequently, followed by mental health, COVID-19, chronic disease, and sexual and reproductive health. YouTube was the most widely used (69.7\% of the participants) platform, followed by Instagram, Facebook, Twitter, and TikTok. The most widely used devices for watching health videos were smartphones, followed by laptops, tablets, desktop computers, and smart TVs. Regarding access, participants browsed health videos in feeds (51.6\%) and “similar videos” recommendations (48.4\%), actively searched for them on social media (51.6\%) or search engines (mostly Google, 41.0\%), or followed recommendations from interpersonal communications with family and friends (26.2\%) and health professionals (22.1\%).} 
  \label{fig:access}
\end{figure}

After \textit{deciding} to watch health videos, participants \textit{access} videos, characterized by cross-topic, cross-device, cross-platform, and cross-channel behaviors. Figure ~\ref{fig:access} showed survey respondents' health video access activity, including access frequency, topics, platforms and devices, and channels. 

\subsubsection{Cross-topic.} Most participants reported consuming diverse topics of health videos because \pquote[P09]{It [health] is a big topic \elide{} mental health, physical health, and all, they are all connected.} Overall, these topics were mostly related to mild conditions. General wellness (e.g., exercise, diet, sleep) was watched the most frequently, followed by mental health, COVID-19, chronic conditions, and sexual and reproductive health. Participants accessed a wide range of topics of health videos frequently with nearly 84\% of the survey respondents (n=121) accessing at least once a week. 

\subsubsection{Cross-device.}
Participants also used multiple devices to access health videos depending on the contexts, as commented by P04, \pquote{between classes, I would use my phone, but if I am doing follow-up exercises, I probably use the tablet or my laptop.} The most frequently used devices were smartphones, followed by laptops, tablets, desktop computers, and smart TVs. 

\subsubsection{Cross-platform.} 86\% survey respondents and 14/18 interviewees accessed health videos through multiple platforms. YouTube was the most widely used (69.7\%) platform, followed by Instagram, Facebook, Twitter, and TikTok. 
We further classified social media as \textit{content communities} and \textit{Social Networking Sites (SNS)} according to \textcite{kaplan_users_2010}. Content communities highlight creating and curating content across broad topics (e.g., YouTube and TikTok), and SNS emphasize building connections and interaction with social relationships (e.g., Facebook and Twitter). We recognized that these two types sometimes overlap because nowadays many platforms have approximately equal emphasis on content and social aspects. For example, Instagram has features of both types that allow users to access a huge number of videos and images while interacting with social relationships.
Some participants directly accessed health videos from SNS or content communities.
It was worth noting that participants also accessed health videos from SNS and then consumed them on content communities. P06 shared “\textit{One of my friends likes to post things from YouTube, Twitch \elide{…} from there, I can trace more videos on YouTube.}”

\subsubsection{Cross-channel.} 
Searching, browsing, and social recommendations are the three main channels of video access.

Participants both actively search for videos as well as passively browse videos, and the two access methods have similar prevalence among the survey respondents. Specifically, participants performed searches both on social media (51.6\%) and search engines (41.0\%). Some participants searched on Google while some participants directly used social media to search for videos, as described by P05 \pquote{I just went to YouTube.}
Participants also passively browsed feeds (51.6\%) and “similar videos” recommendations (48.4\%) to access videos. Passive encounters were more common when participants did not have a clear need such as during leisure time, \pquote[P17]{I just keep scrolling and watching videos on TikTok.} 

Social recommendations are suggestions from interpersonal communications with family and friends (26.2\%) and health professionals (22.1\%). Around one-fourth of the participants reported accessing videos via social recommendations.
Social relationships recommended videos based on their knowledge of the participants’ interests, as P07 commented “\textit{They are just like, try this out, maybe you like it. I think I told them to help me learn certain things, and they just share when they find something beneficial for me.}” 
The recommendations could be a platform, an account, or links to videos. For example, P08’s friends directed him to YouTube to search for videos related to coping with depression; P17’s doctor recommended to him a YouTube account about diabetes; P18’s doctor sent her several links to videos about exercises for Parkinson's. 

\subsection{Watching videos}
Once having access to videos, participants watch them. Watching videos involves multiple activities, including processing information, taking notes, and experiencing emotions. 

\subsubsection{Processing information} Participants considered processing information in the video format easy, efficient, and effective. 

\textbf{Understanding information with greater clarity.} Combining visual demonstrations and oral explanations, videos convey information in a clear manner, particularly for learning procedural information about how to do something. Videos also made it easier for visual learners to understand information better. For example, P07 said“\textit{Visually seeing diagrams and pictures helps to understand. I don't like reading because I get lost in the words sometimes.}”

\textbf{Obtaining information with higher efficiency.} Videos were perceived to be more efficient for obtaining information in terms of time and cognitive cost. For example, P14 commented “\textit{[Video] is one of the most efficient ways to get information because they are usually pretty short. They're really efficient versus long articles, which can be really boring with lots of ads \elide{or pop-ups that can be really hard to read.}}” The fast pace of videos also contributed to the perceived efficiency like P17 thought “\textit{Dr. Schmidt speaks so fast that you get so much content in a minute.}”

\textbf{Getting longer attention span.} Many participants compared watching videos with reading books and concluded that videos are effective in keeping their attention. For example, P11 commented “\textit{[Videos] are able to deliver information which is palatable to my brain. Otherwise, I would have to be in some libraries \elide{when I got a mind reading…} with my attention span, I’d constantly scatterbrained all over the place.}” P10 compared videos with texts and images and commented that “\textit{[On Instagram], maybe people don't watch the whole video but they probably will stay engaged to it longer \elide{than a bunch of text, images.}}” 

\subsubsection{Taking notes}
\begin{figure}[h]
  \centering
  \includegraphics[width=0.4\linewidth]{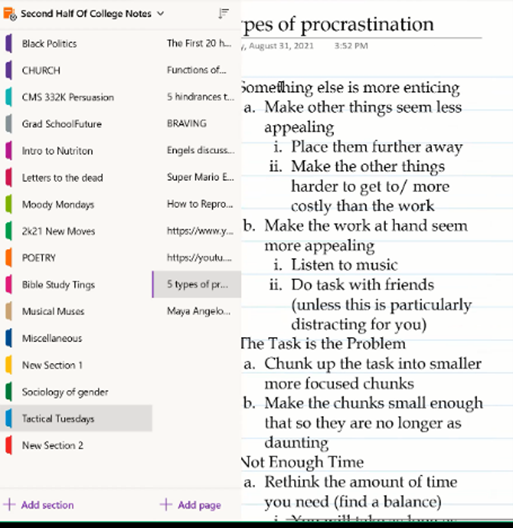}
  \caption{Notes taken by P02 while watching health videos}
  \label{fig:notes}
  \Description{Notes taken by P02 while watching health videos}
\end{figure}

Some participants reported taking notes while watching videos, particularly those with a purpose to learn about a subject. For example, P02, to open a podcast channel to propagate health knowledge about mental health, said “\textit{I will watch mental health-related videos and take notes to strengthen the learning effectiveness.}” Figure ~\ref{fig:notes} showed a screenshot of her notes. Some participants expressed the need to take notes but did not resort to external tools. For example, P07 said she took "\textit{mental notes}" and P13 tried to "\textit{memorize}" the content.

\subsubsection{Experiencing emotions} Participants reported emotional arousal while watching videos, including a stronger sense of connection with creators, a sense of realness, and a sense of fun.

\textbf{Connecting with content creators.} The perceived “\textit{face-to-face}” communication and the creators' real voices and performances connect viewers to creators. For example, P07 commented “\textit{I prefer them [creators] appearing [in videos] because they are facing you directly like they are having a conversation with you. That makes you connect to the information that they are trying to portray.}” P01 said that “\textit{the way that she was talking about her experience, how she suffered, and nothing was working. That's like, she knows what I'm talking about and I know she's talking about.}”

\textbf{Getting immersive experience.} Some viewers developed a sense of realness by being immersed in the contexts, environments, and lives of the vloggers. P03 commented on her experience of watching first-person shooting vlogs of vegans “\textit{They follow them throughout the day, show what kind of stuff they do, so you know they cook this kind of stuff in their breakfast. You see through their eyes, the process, how they go shopping, how they cook, how they supplement.}” P05 said “\textit{I think being able to relate to somebody rather than just reading plain text of an article was helpful. Videos talking about somebody who had anxiety, \elide{walked through it, and they had some coping methods… } the ethos, and rhetoric. [I felt] like they went through it and they were able to thrive.}”

\textbf{Exploring diverse genres of content.} Some viewers engaged with videos' versatile genres, such as cartoons, imitations, commentary, interviews, talks, and vlogs, which afford diverse and novel styles and make the content more interesting, engaging, and fun to watch. For example, P10 remarked on a doctor influencer whose videos featured imitation: “\textit{He's always playing every character. He'll be this doctor, then he puts on a wig to be a nurse. It's all script-based to get you to think about some health topics that happen in the hospital, and they're really funny. I was like, oh my gosh he's hilarious, and I followed him.}” P10 further commented, “\textit{I think these videos are interesting because every video is different. It's not like the same setup or template. Everyone is new and exciting.}”

\subsection{Evaluating videos}
Participants form evaluations of videos based on their watching and engagement with videos. They assessed the credibility of videos from three aspects: previews on platforms, video content, and video creators. We extracted the specific indicators they used in Tab \ref{tab:evaluation}.

\begin{table*}[h]
\caption{Evaluation of Video Credibility}
\label{tab:evaluation}
\scriptsize
\begin{tblr}{
  width = 0.9\textwidth,
  colspec = {|Q[l,0.6]|Q[l,0.7]|Q[l,3.5]|}, 
  hlines,
  row{1} = {font=\bfseries,c,gray9},
}
                    &  Indicators                          & Definition    \\
\SetCell[r=3]{m}{\textbf{Platform previews}} & Covers & Covers with more information is more credible  \\
& Metrics & Videos with more view numbers and like numbers are more credible \\
& Account type & Organizational and governmental accounts are more credible \\
\SetCell[r=2]{m}{\textbf{Video content}} & Edits & Careful edits such as the design of visuals, voices, and background music indicate higher credibility \\
& Humors & Humorous elements are perceived to be more credible \\
\SetCell[r=4]{m}{\textbf{Video creators}} & Appearance  & Seeing the face of the presenter or creator is perceived more credible \\
& Credentials & Professional-created videos are more credible\\   
& Tones & Enthusiastic tones suggest stronger motivation and higher credibility \\
& Personality & Creators with a stronger personality (e.g., vocal, rigorous) are more credible
\end{tblr}%
\end{table*}

\paragraph{Platform previews}
When searching on Google or YouTube and selecting videos to view from the results list, participants would form an initial evaluation based on video features such as ranking, title, cover, and metrics. For example, P01 said, \pquote{I look at the video cover, the title of the video, the account name, and how many views, like a million views is probably enough for me.} Participants acknowledged that they tended to click videos with visually appealing or "\textit{attention-grabbing}" covers.
Five participants watched health videos produced by non-profit organizations, patient communities, and medical students and professionals. For example, P07, P11, P14, and P17 all watched mental health videos on YouTube from Psych2Go, an organizational account, for its professional and comprehensible content.

\paragraph{Video content} 
Viewers also valued video with careful edits, which entailed the selection of appropriate background music, voices, and images. A human voice was preferred to a machine-generated voice when explaining health knowledge. For example, P06 said, "\textit{I like a real person talking because narrated voice helps me focus more and understand more.}"
Humorous elements and entertainment genres were generally preferred by participants. P04 remarked "\textit{I trust them [humorous videos] more because I like the way they deliver the workout. Some YouTubers are so boring.}"

\paragraph{Video creators} 
15 out of the 18 participants watched user-created health videos. Many video creators appeared and talked in the video. Some participants trusted video creators who intensively self-disclose and show strong personalities. 
The Credential of the creators was not limited to a professional certificate in health-related areas but included shared experiences in a broader sense.

Some participants were in favor of passionate and enthusiastic tones as reflected in their talking, and they interpreted such tones as indicators of good intention. For example, P13 liked a YouTuber who "\textit{has a face very enthusiastic about sharing the information,}" and P02 trusted a YouTuber because "\textit{He's a very excitable and vocal person that you can feel something he says it. He seems earnest in his pursuit.}" 

The strong personality of the creator also plays a vital position in reinforcing participants' trustworthiness. Participants seemed to have diverse expectations of and preferences for health video creators regarding different topics.
Some participants trusted creators who were precise and rigorous. P12 trusted a Korean vlogger who tests, presents, and explains the effects of various skin care products. She commented that she"\textit{is not necessarily professional but definitely does lots of research}" and "\textit{always shows the factual information... brings in skincare curator sometimes.}"
However, four participants were attracted to creators with distinct personalities compared with others. P02 expressed appreciation for a creator who objected to and satirized fitness influencers who promote quick weight loss methods. She interpreted the aggressive words and sometimes exaggerated facial expressions of this creator as signs of "\textit{authenticity}" because "\textit{he doesn't care about whether you like him or not and he is just saying how he feels.}" 

\subsection{Using health videos}
Participants reported three main ways of \textit{using} health videos: assisting problem-solving, guiding long-term behavioral changes, regulating emotions alone with positive videos, and managing content.
13 of the 18 participants successfully achieved at least one goal, and they accredited the achievement to health videos on social media, reporting that 50\% to 90\% of the success was due to videos.

\subsubsection{Assisting problem-solving}
When participants searched health videos with clear and many times urgent problems, such as treating eye problems, making healthy meals, and exercising, they typically used the videos while or immediately after watching them. For example, P01 tried six methods from YouTube videos after doctors failed to treat his Chalazion, including washing his eyes, using vinegar, and massage. He was anxious to be cured, as he commented “\textit{I wanted to try anything and everything to pop it, although you're not exactly sure how it works\elide{. Maybe it's antibacterial or it's anti-inflammatory.}}”

\subsubsection{Guiding long-term behavioral changes} 
\begin{figure}[h]
\centering
  \includegraphics[width=.4\linewidth]{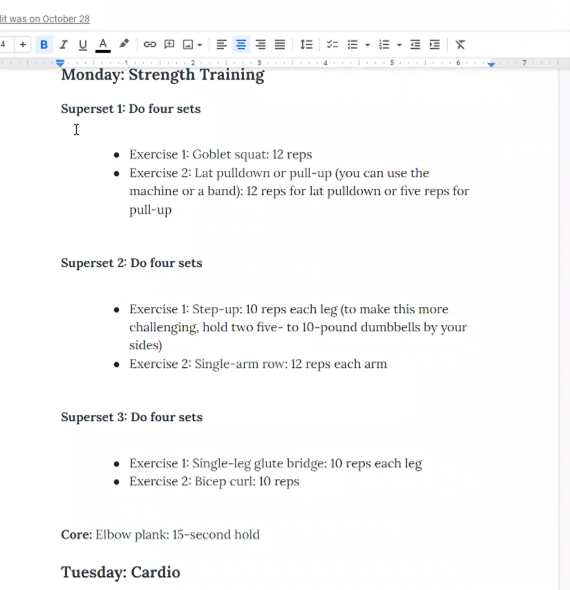}
  \caption{Plans made by P10 to guide weight loss}
  \label{fig: plans}
\end{figure}

Participants also used health videos to guide longer-term behavior changes. By consistently following health videos, P08 overcame symptoms of depression and four participants lost weight. To realize the weight loss goal, P10 applied the knowledge that she learned from a variety of health videos and made an exercise plan (as shown in Fig \ref{fig: plans}), specifying a workout routine over a week. P18 reported performing two sets of Parkinson's wellness recovery exercises every day following videos from YouTube to get a “\textit{stronger body}” and a “\textit{better mood.}”

\subsubsection{Regulating emotions alone with positive videos}
Ten participants shared experiences of successfully getting emotional support to varying extents. Being able to regulate emotions alone was appreciated by participants because health videos afforded privacy, availability, and easiness. Some participants found videos to be more useful than actual people. P16 said “\textit{If it's a physical problem, like pains, obviously I'll go to a doctor. But if it's just stressful. I would turn to videos because it is more of an emotional problem… it's quite hard to have consultations with friends and family about it.} 

Participants selected positive videos for use to inspire and encourage positive behaviors. For example, P11 said "\textit{They can get rid of carbs. That would be a trigger for my mind. And that makes me know that I can [do it], too.}" P16 also said, "\textit{If I'm struggling with pain or emotional aspects of living with it [the disease], I'll search out positive videos to help me with dealing with pain.}"

\subsubsection{Managing content}
\begin{figure} [h]
\centering
\includegraphics[width=.4\linewidth]{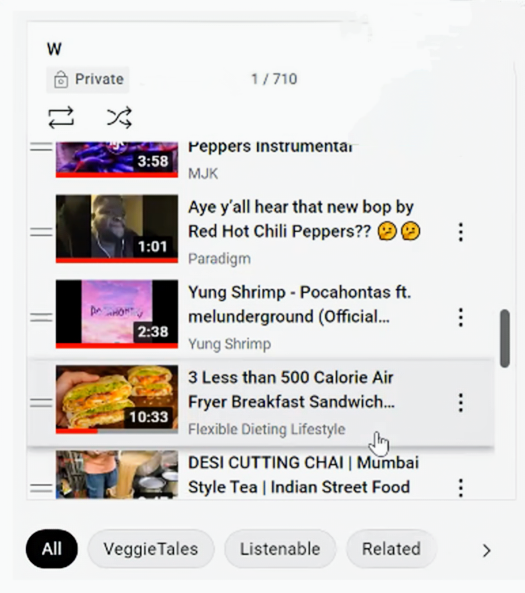}
\caption{P11's YouTube folder to collect videos}
\label{fig:management}
\end{figure}

Managing videos and/or associated information artifacts (e.g., notes and links) is an integral part of using video information for some participants, particularly when they plan to reuse the videos. For example, P18 saved the email sent by her friend which contained the YouTube Parkinson's recovery exercise videos, and pulled out this email daily to do follow-up exercises. Some participants manually created or designated one folder to “star” or “collect” videos. The fact that there were no functions to support video management led some participants to mix all videos into one folder. For example, P11 had a playlist on YouTube called “\textit{'W' where I just throw a whole bunch of videos. If I'm ever on a diet and ideas, that's still good for me to keep,}" but it troubled him that "\textit{There are about 700 videos as you can see (Fig ~\ref{fig:management}), lots of them aren't health-related. I have to go back through [this folder] to see whatever I have watched.}” 

\subsection{Iterations and dynamics among the five activities in the video consumption model}
Participants' health video consumption is dynamic and iterative in nature. As illustrated in Fig \ref{fig:model}, participants may skip one activity or return to an earlier activity in the consumption process. 
For example, after \textit{deciding} to watch videos to fulfill certain motivations (e.g., entertainment), some participants iterated the \textit{accessing}, \textit{watching}, and \textit{evaluating} activities. After several iterations (i.e., consuming some health videos), they developed new motivations (e.g., health improvement) and started to \textit{use} the health videos to guide behavioral changes. Later, they might also change access channels based on the evaluation of videos across platforms.

\begin{figure}[h]
  \centering
  \includegraphics[width=0.8\linewidth]{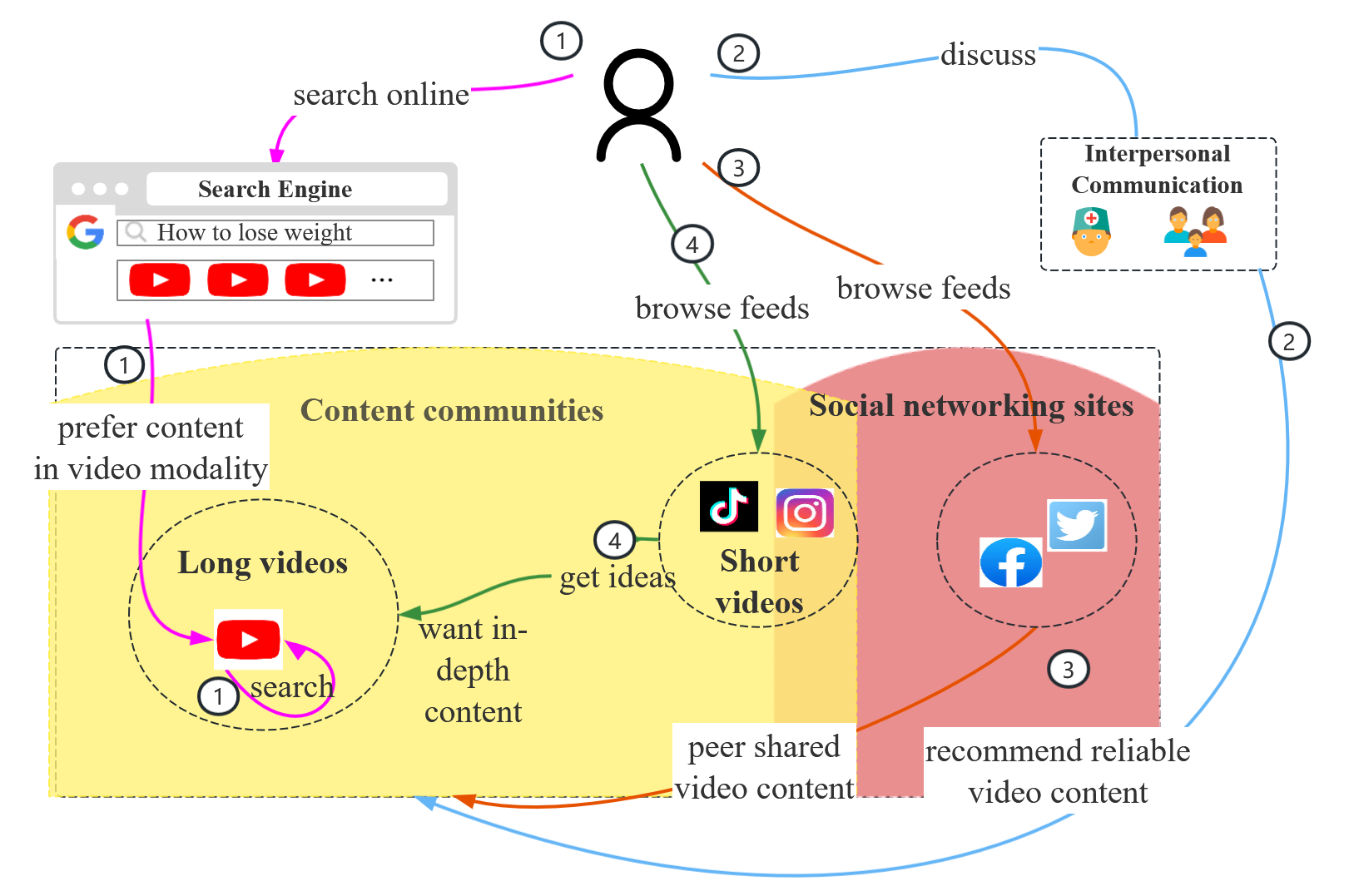}
  \caption{Dynamics and Iterations in the Five Activities of Health Video Consumption}
  \Description{Ecosystem of accessing health videos on social media. We classified social media as content communities and social networking sites according to \cite{kaplan_users_2010}, with content communities highlighting the content aspects (e.g., YouTube and TikTok), and social networking sites emphasizing social aspects (e.g., Facebook and Twitter).}
  \label{fig:ecosystem}
\end{figure}

We found that these dynamic consumer activities and technology environments collectively co-constructed the socio-technological ecosystem (shown in Fig \ref{fig:ecosystem}). The ecosystem comprises social media, including content communities and Social Networking Sites (SNS), search engines, and participants' social relationships. 
In this dynamic and iterative process, we observed several tendencies of health video consumption, including \textbf{seeking better accessibility and higher reliability, and forming stronger motivation}.

\subsubsection{Better accessibility: Search engine to social media}
Directly accessing videos from social media offers better accessibility. For example, P01 started searching for Chalazion-related information on Google and selected YouTube videos that were ranked at top positions on the result page as illustrated by \textcircled{1}. After several iterations of redirection from Google to YouTube, he experienced more convenience on YouTube. He \textit{accessed} and \textit{watched} many recommended videos and he could easily modify queries on YouTube for refined search results. He described \pquote{I actually just went to YouTube and searched the topic rather than on Google.}

\subsubsection{Higher reliability: Social recommendations and longer videos}
Participants relied more on social recommendations after iterative consumption of health videos from various channels, especially if they planned to \textit{use} health videos. The social recommendations could happen at offline interpersonal communications (\textcircled{2}) as well as interactions on SNS (\textcircled{3}). Half of the participants \textit{watched} and \textit{used} health videos recommended by their social relationships. These videos were almost always relevant and credible because they were recommended by doctors, family, and friends who knew better about their needs. For example, P16 immediately trusted the YouTube account "DIABETES CODE" recommended by his doctor and \textit{used} it to learn how to prevent diabetes. 
P17 and P18 were members of patient communities and they \textit{watched} health videos produced by these communities.

Participants also preferred longer and more in-depth videos for \textit{use} after several iterations as illustrated by \textcircled{4}. Many commented that YouTube is the most suitable platform for \textit{accessing} and \textit{watching} in-depth health videos since videos on YouTube were longer and more “\textit{reliable} and “\textit{credible}.”

\subsubsection{Stronger motivations: entertainment to learning and behavioral changes}
Motivations that drive participants to \textit{decide} to consume health videos may develop over the iterative consumption process. Sometimes seeking social awareness and entertainment might transform into learning and even trigger behavior changes, reflecting that many participants got new ideas about lifestyles from short video platforms and went to YouTube to learn more (illustrated by \textcircled{4}).
Some participants strategically \textit{accessed} short videos to encounter \enquote{new things} or look for \enquote{ideas} (e.g., food cleansing, keto diet) and get motivated (e.g., gym hack) from the trending videos. Although \pquote[P03]{TikTok and Instagram are not the most reliable, but they're very quick}. They turned to YouTube videos to support \textit{use} activities. P17 gave an example “\textit{I was scrolling on TikTok and something popped up, it’s about hypertension and I found it interesting. But it did not go through deep content, so I went to YouTube.}”

\section{RQ2: Challenges of Health Video Consumption on Social Media}
Participants encountered challenges in each health video consumption activity. Small barriers at earlier activities can accumulate into big challenges later (e.g., \textit{using}). We summarized challenges associated with each activity in Tab \ref{tab:challenges} for clarity, but it should be noted that some problems in downstream activities may be rooted in challenges from upstream activities.


\begin{table*}[h]
\caption{Challenges of Health Video Consumption on social media}
\label{tab:challenges}
\footnotesize
\begin{tblr}{
  width = \textwidth,
  colspec = {|Q[r,0.4]|Q[l,0.7]|Q[l,3]|}, 
  hlines,
  row{1} = {font=\bfseries,c,gray9},
}
Activities & Challenges & Example quotes \\ 
\SetCell[r=1]{m}{Deciding} & High expectations &  “for some people, they are looking for overnight weight loss and they might end up falling victims of the viral videos in weight loss.” (P02) \\ 
\SetCell[r=2]{m}{Accessing} & Lacking video content retrieval & "it's usually not the first thing that pops up on your result, and it is frustrating because I have to dig deeper for resources that I am looking for" (P03) \\
 & Low professionality & “there are so many people talking while they are not qualified to talk” (P14) \\
\SetCell[r=2]{m}{Watching} & Overwhelming recommendations & “For me, it’s very transactional and I don’t want them full of my feeds” (P01) \\
 & Negative emotions & "I don't want to see these recommendations a lot, because I can easily empathize with their sufferings." (P16) \\
\SetCell[r=3]{m}{Evaluating}  & Commercial intentions & “a lot of the market profit off the fact that people do not know the correct things and keep making mistakes.” (P02) \\
& Contradictory content & "There are too many resources give you contradictory suggestions." (P09) \\
 & Low literacy & "the green check on YouTube means this person is verified as a doctor or some professionals" (P15) \\
\SetCell[r=5]{m}{Using} & Personalization & "It's really hard because everyone's metabolism and how their body works is different, so you have to really learn how your body works"(P12) \\
 & Failure & "After a while, I got frustrated about not seeing the results you want, especially when I was working out so much and I don't understand." (P12) \\
 & Integration with tracking apps & “but I didn't write it down just because it's hard for me to like completely type out but that like I do myself." (P10) \\
 & Lack of feedback & “I don’t know if I am making progress”. (P18) \\
 & Content management & "This is where I just throwing a whole bunch of videos right here, so I have like 700 videos as you can see." (P11) \\ 

\end{tblr}
\end{table*}
\paragraph{The over-expectation of health videos.} Four participants failed to accomplish their health goals, indicating a misalignment between \textit{deciding} and \textit{using} activities. Some participants expected to use health videos on social media as a substitution for professional treatment. Such expectations are risky and often lead to disappointments. For example, P11 shared her frustration when health videos failed to help her, \pquote{it was so discouraging. I got very frustrated because I felt like I was trying everything but nothing was working.}

\paragraph{Demanding efforts to find relevant videos.} The lack of sophisticated content filters and in-video retrieval demands extra time and substantial effort for participants to go through videos. In some cases, it only led to a finding that some videos are irrelevant. Professionality of health videos is a dimension of relevance for some participants. Many platforms only have limited search functions such as ranking and filtering videos by metrics that are unhelpful for determining content relevance. Participants had to rely on the title and cover to decide whether or not to watch. Some platforms did not support search as P08 commented, "\textit{[Facebook] was not designed for search, it will return a list of people or groups}."

\paragraph{The uncontrollable algorithms.} Participants were exposed to overwhelming recommendations and potential negative emotions that they did not expect. Although participants acknowledged that recommendations could be helpful, sometimes participants felt they were \enquote{annoying}. Many participants tried to control their interactions with videos to avoid overload by using strategies such as intentionally not following or subscribing to relevant accounts, particularly when they had problem-solving goals. P16 was a highly "\textit{sensitive}" that \textit{"I don't want to see these recommendations a lot, because I can easily empathize with their sufferings."}

\paragraph{Multifaceted difficulties of making an assessment.} Evaluation is a common challenge in health information consumption. Participants considered the quality of health videos on social media varying, causing difficulties in evaluating video credibility. 
Some participants attributed this challenge to the fact that publishing health-related topics on social media does not require a certificate. With the user-generated health videos trending, the enhanced sense of closeness and interactivity curated by creators' communication techniques blur the boundary between credibility and engagement.
Participants also cited commercial or for-profit interests involved in some videos. 
Insufficient knowledge about platforms is another reason; for instance, P17 considered videos on YouTube more legitimate because they could be retrieved by Google.

\paragraph{Obstacles of fitting information to personal needs.}
Despite being relevant and helpful, solutions offered in health videos often could not be directly applied to the participants and required personalization. P03 was also bothered by an idiosyncratic joint problem that “\textit{it's like a black cartilage film… so my bones are closer together in there and have less padding}.” Thus, there were no ready-made solutions for her to refer to. These difficulties sometimes originate from the accessing activity, where the participants could not find the most relevant videos that suit their needs; rather, they had to transform the knowledge learned from other videos into self-experimented solutions. Difficulties in forming personalized solutions can lead to failure to use health videos to guide behavior changes. 

\paragraph{Low sustainability and efficiency for behavioral changes.} Even if participants have a plan for behavioral changes, solely using health videos is not sustainable or efficient. Behavioral changes go hand-in-hand with self-tracking, feedback, learning, and reflection spanning a long time. However, video platforms can not be linked with other self-tracking apps. P12 had to monitor his calorie consumption by manual calculations using the formula he learned from YouTube videos to cope with prediabetes.
Participants expressed the frustration of following the health videos but could not get any feedback. P18 exercised every day following workout videos, but lamented that \textit{“I don’t know if I am making progress.”} 
Further, video platforms only support the collection of videos but have less support for management functions, such as recommending classifications of content.

\section{Discussion}
Most existing studies about users' consumption of health information on social media mainly centered around text information. In response to the fast surging of health videos on social media, we interviewed and surveyed consumers about their health video consumption practices on social media with two purposes: to enhance the conceptual understanding of health video consumption on social media, and to shed light on implications for designing more effective socio-technical environment to realize the potential of videos in promoting public health and wellness and supporting individuals' behavioral changes. In this section, we first characterized video consumption practices based on the health video consumption model that we developed and then discussed design implications. 

\subsection{Characterizing video consumption}

\subsubsection{Deciding to watch videos: Social awareness and entertainment as new motivations} 
The Use and Gratification Theory (UGT) postulates that people intentionally (rather than passively) select media to satisfy certain social and psychological needs \cite{katz1973uses}. The deciding activity, particularly the motivations outlined in our model, corresponds to the gratifications in the UGT. It is not surprising that consumers chose to watch videos to improve health \cite{huh_collaborative_2012, mamykina_collective_2015}, learn health knowledge \cite{de_choudhury_seeking_2014, hartzler_evaluating_2014}, and regulate emotions \cite{coulson_receiving_2005, grimes_eatwell_2008}. These motivations or gratifications were well documented in research on consumers' health information-seeking behavior on online health communities and social media platforms based primarily on analyzing consumer health communication in text format \cite{zhang2013facebook}. Our study indicated that they also apply to videos. 
 
Our results also identified two additional motivations for accessing health videos compared with text: obtaining social awareness of important others' interests and seeking entertainment. This result is related to the affordances of social media platforms where awareness of one's social relationships is possible and the affordances of video as an information-rich media \cite{daft1986organizational} that can provide a more visual and immersive experience. It's worth noticing that motivations may change as consumers iteratively interact with various health videos. Many participants' curiosity transformed into serious learning, motivating them to seek more reliable information from long videos. In our study designated to health topics, a number of participants were well motivated to learn and future research can investigate how to trigger entertainment for learning purposes. However, in daily video consumption for entertainment, this behavior is probably less prominent. 

\subsubsection{Accessing videos: Cross-platform information-seeking and encountering}
Research in HCI and information-seeking and retrieval typically focuses on user behaviors on one information system or platform. Researchers have started to note many information behaviors, such as sharing everyday activities \cite{zhao_social_2016} and disclosing sensitive personal information \cite{pyle_lgbtq_2021} and identity \cite{devito_too_2018} are cross-platform. The socio-technical ecosystem of health video access consists of search engines, interpersonal communication, and different types of social media (i.e., content communities versus social networking sites and long-video platforms versus short-video platforms). The multiple access pathways that we identified indicated that consumers' health video access behavior is also cross-platform.

Another noteworthy finding of the study is that actively seeking or searching for health videos and passively encountering or receiving health videos on social media share equal prevalence in enabling access to health videos. This phenomenon was also reported by Sun's study on college students' social media consumption behavior \cite{sun2022losing}. In today's technology environment, browsing and encountering information are playing an increasingly important role as the active seeking of video information is indicative of users' interests and thus leads social media platforms to push more relevant content to viewers. As a result, the information environment is becoming increasingly responsive and proactive such that users are subtly coerced to become more passive in information seeking and consumption, raising concerns about the power of social media algorithms and algorithmic biases. 

\subsubsection{Watching videos: Cognitive clarity, emotional engagement, and genre-enabled sense of interestingness and novelty}

We added to the existing knowledge that compared to text, videos allow consumers to more effectively process information. Videos provide a powerful platform for information understanding, excelling in their ability to convey multimodal content \cite{mayer_multimedia_2002}. They effectively combine visuals, audio, and text, making complex ideas more accessible and engaging, while simultaneously offering efficiency through succinct communication. 
Moreover, videos enhance engagement and memory by capturing tone and facial expressions. Videos are also readily accessible on mobile devices. 
The preference between text and video depends on factors like audience, content nature, and communication objectives. Some participants, particularly those with serious learning goals, transform visual information into text notes.

Previous literature reported that online videos afford immersion, interactivity, and intimacy \cite{harris_young_2021, haimson_what_2017} and thus can enrich personal and contextual disclosure \cite{liu_health_2013}, cultivate a sense of closeness \cite{southerton2021research}, and a sense of community \cite{harris_mixed_2019}. Our participants also reported similar effects, pointing out that videos cultivate a strong sense of connection with creators, a sense of realness, and a sense of fun. Specifically, we identified that directly talking to cameras and \enquote{in situ} shooting that showed details of experiences are essential to cultivate effective social support without requiring direct social interactions.

\subsubsection{Evaluating videos: Importance of visual designs and creator features}
The quality of health videos on social media is problematic \cite{gabarron2013identifying, krakowiak_youtube_2021, sledzinska_quality_2021}, but few studies examined how viewers evaluate online videos \cite{gamage2022designing}. Generative AI created videos and deepfake technologies \cite{verma2022artifice} can further complicate the social media ecosystem and environment, raising challenges in people's video evaluation \cite{niu2023building}. To our knowledge, this is the first study providing insights about how people evaluate health videos on social media. We demonstrated that participants evaluated the access channel (i.e., how they get this information), video content, creators, and some platform information such as ranking and user engagement statistics. We also found that assessing videos is a complex process involving not only credibility but also other dimensions such as trustworthiness, interestingness, and visual attractiveness \cite{sun2019consumer,liu2023consumer}. These findings are consistent with studies focusing on users' assessment of mainly textual health information \cite{liu2021linguistic, zhang2023design}. Nevertheless, we found that, compared to text information, users relied more on affective measures to evaluate videos, such as creators' features, including their appearance, facial expressions, voices, and personalities, whether or not they disclose personal information, and whether or not they sound enthusiastic and authentic. These criteria seemed to be related to parasocial interactions (PSI) \cite{horton_mass_1956} viewers experience during video watching, where viewers develop one-sided intimacy and emotional reliance towards the creators \cite{niu_stayhome_2021}.

\subsubsection{Using videos: Behavioral changes and personalization practices}
Consistent with prior research on consumer health information behavior, our study found that consumers use health videos to learn about health topics, regulate emotions, and guide behaviors \cite{lambert2007health}. We extended the understanding by revealing that consumers often go through a personalization process to fit the information to their physical or life constraints. This process requires additional work such as experimentation, data collection (e.g., through self-tracking), reflection, and planning. We also found that some consumers exerted efforts to manage the videos they found for future access and use. 

\subsection{Design Implications}
\paragraph{Creating engaging videos to motivate access.} The existence of different motivations for accessing videos suggests that health educators, particularly healthcare providers, government agencies, and non-profit health organizations, should consider creating video collections that have multiple genres that use different styles to convey similar health content. Generative AI has been increasingly used by video content creators \cite{lyu2024preliminary} to elicit ideas, write scripts, and even impersonate characters to accommodate different user needs. Health educators should strategically use such tools to produce more engaging content and encourage user access \cite{song2022serious}.

\paragraph{Broadening access channels.} The socio-technical ecosystem and multiple pathways of video access suggest that health educators should work to provide multiple ways to promote access to their videos. First, they should post their videos on social media to allow them to be picked up by consumers while browsing their feeds. Second, health educators should distribute their videos across platforms and tailor the content to suit different platforms. For example, they may create short videos to put on short video sites like TikTok or YouTube Short to serve as a lead to or preview of more substantive videos on YouTube or their own websites to increase the chance that the substantive educational videos will be accessed. Third, health educators should also make use of the power of interpersonal communications by distributing information directly to their customers through methods such as emails or patient portals and making the content easily sharable to family and friends.

\paragraph{Supporting learning.} Video designers may create innovative ways to provide structured information, such as table of contents, section headings, and summaries, in videos to better support some consumers' needs for taking notes about the content. Social media platforms may also provide tools to support note-taking, such as generic note-taking functions and functions that allow viewers to extract portions of the videos or transcripts and insert them into their text notes.

\paragraph{Identifying high-quality videos.} Prior studies found that low-quality and misleading health videos on social media had negative impacts such as self-doubt, anxiety, and disappointment \cite{jordan_can_2021}. Common intervention strategies to support consumers' evaluation of online health information include educational programs, algorithms to automatically identify high-quality content, and interactive interfaces giving real-time judgment assistance \cite{song_interventions_2021}. Considering the sheer amount of health videos on social media, deploying algorithms such as deep learning \cite{al2017using}, to be combined with selective manual assessment, is needed to assess video quality and promote high-quality videos. Interfaces could be developed to assist viewers as well, such as making visible some effective quality indicators, such as video sources. Given a wider use of affective criteria to evaluate videos in comparison to text, educational materials should also be developed to help people be aware of their behaviors and cautious about potential biases involved, particularly for groups with comparatively lower technology proficiency or information and health literacy. 

\paragraph{Assisting consumers' use of videos.} (1) Supporting personalization. Functions, such as templates to support the creation of exercise and diet plans (for those who watch such videos) and structured forms to support the recording of self-experimentation, could be designed to support self-experimentation and planning. Content creators may include in videos how viewers may observe progress, such as tracking and reflecting on certain parameters. (2) Supporting content management. Platforms could offer functions to help manage videos for re-finding and reuse, such as creating categories and organizing collected content automatically after users "star" the content, allowing viewers to indicate the importance of a video and rank videos by importance, and allowing viewers to tag and annotate saved videos. In addition, cross-platform content management should be supported considering consumers' access and use of videos are cross-platform \cite{devito_too_2018}.

\section{Conclusion}
Health videos on social media are popular and widely consumed, but the understanding of users' practices of consuming health videos on social media is limited. We interviewed and surveyed users and identified five main activities involved in their health video consumption practices: deciding, accessing, watching, evaluating, and using. We constructed a descriptive model of video consumption on social media to demonstrate the general process of health video consumption on social media, the socio-technical systems that enable and constrain the process, and the cognitive, emotional, and behavioral actions involved. Future research may focus on detailing specific components or processes in the model or relationships between different activities. For example, how motivations, content, genre, and video affordances impact consumers' watching and evaluation of videos; how to support consumers to integrate information from multiple videos and cross multiple platforms to enable better personal health planning. Our study also generated a number of design implications to enhance the video information ecosystem on social media to better serve consumers' health-related needs, such as diversifying genres in delivering health content to meet differing motivations and video-watching preferences and developing tools to support note-taking during video watching. 

\section{Acknowledgments}
This research was supported by the University of Texas at Austin, School of Information, John P. Commons Teaching Fellowship.

\bibliographystyle{ACM-Reference-Format}
\bibliography{bib}

\appendix
\section{Survey Details}
\subsection{Survey Design}
The survey instrument consisted of three components: (1) respondents’ video-watching practices on social media, including platforms and devices used, frequency, topics of interest, video access, watching, sharing, and management behaviors, and the use of information obtained from videos; (2) respondents’ motivations for watching health videos on social media. The motivation items were derived from a review of prior literature and a preliminary analysis of 12 interviews that had been conducted when the survey was being designed; and (3) demographic information, such as age, gender, and education. 
We tested the initial survey instrument with eight participants in two rounds. Feedback included reducing the number of options in multiple-choice questions, clarifying the wording of questions and answers, and improving the mobile display interface, which was incorporated into these iterations. To enhance data quality, we also added one attention check question, asking respondents to select the number of options they checked for a prior multiple-choice question \cite{august_pay_2019, shamon2019attention}. The final survey contained 33 questions and was launched on Qualtrics.

\subsection{Survey Data Exclusion}
Before the data cleaning, there were 147 from MTurk, 41 from the university listserv, and 360 from Reddit.
We excluded responses that met any of the following criteria: (1) incomplete; (2) completed in less than three minutes; (3) failed the attention check question; and (4) gave irrelevant responses to the open-ended questions. We also excluded responses submitted during the same short time period and had the same email address patterns (we asked participants’ emails to enter a random drawing of a \$30 Amazon gift card). 
There could be biases when determining whether a survey respondent should be included. Survey responses collected via Reddit and MTurk demonstrated features that could be related to bots, such as the number of responses submitted during a short period and similar email patterns (e.g., name plus five random digits followed by @ and the domain name). The exclusion of such data may have systematically excluded certain kinds of participants.
\label{suveydetails}
\end{document}